\documentclass[conference, a4paper, 10pt]{IEEEtran}
\pdfoutput=1

\usepackage[english]{babel}
\let\labelindent\relax
\usepackage{enumitem}

\usepackage{color}                                          % Enable color support (i.e. for links)
\usepackage[pdftex]{graphicx}                                 % Use images
\usepackage[nolist]{acronym}                         		  % Enable acronyms
\usepackage{xspace}                                           % Dynamic space in \newcommand
\definecolor{Darkblue}{rgb}{0,0,.5}

\newcommand{\bibnl}{\hskip 1em plus 0.5em minus 0.4em\relax}

\newcommand{\knuthinterval}{\ldotp\ldotp}

\newcommand{\om}{OMNeT++\xspace}
\newcommand{\ether}{Ethernet\xspace}

\newcommand{\sync}{\texttt{Sync}\xspace}

\newcommand{\dreq}{\texttt{Delay\_Requ}\xspace}
\newcommand{\dresp}{\texttt{Delay\_Resp}\xspace}

\newcommand{\lpln}{\emph{LibPLN}\xspace}
\newcommand{\lptp}{\emph{LibPTP}\xspace}

\usepackage[
   pdfauthor={Wolfgang Wallner},
   pdftitle={Simulation of the IEEE 1588 Precision Time Protocol in OMNeT++},
   colorlinks=true,                            % Links are marked as colored text, not colored box.
   linkcolor=Darkblue,                         % The color for in-document links (e.g. in the table of contents).
   citecolor=Darkblue,                         % The color for bibliographic citations.
   urlcolor=Darkblue,                          % The color for hyperlinks to the Net.
   pdftex
]{hyperref}

\addto\extrasenglish{%

}

\begin{document}

% =======================================================================
%     DOCUMENT PROPERTIES
% =======================================================================
\title{Simulation of the IEEE~1588 Precision Time Protocol in OMNeT++}

\author{\IEEEauthorblockN{Wolfgang Wallner}
\IEEEauthorblockA{Vienna University of Technology\\
Email: wolfgang-wallner@gmx.at}
}

%\pdfinfo{
   %/Author (Wolfgang Wallner)
   %/Title  (Simulation of the IEEE 1588 Precision Time Protocol in OMNeT++)
   %/CreationDate (\date)
   %/Subject (PDFLaTeX)
   %/Keywords (PTP, IEEE1588, Simulation, Powerlaw Noise)
%}

% =======================================================================
%    TITLEPAGE
% % =======================================================================
\maketitle

% =======================================================================
%    ABSTRACT
% =======================================================================
\begin{abstract}
Real-time systems rely on a distributed global time base.
As any physical clock device suffers from noise, it is necessary to provide some kind of clock synchronization to establish such a global time base.
Different clock synchronization methods have been invented for individual application domains.
The \acf{PTP}, which is specified in IEEE~1588, is another interesting option.
It targets local networks, where it is acceptable to assume small amounts of hardware support, and promises sub-microsecond precision.
PTP provides many different implementation and configuration options, and thus the \ac{DSE} is challenging.
In this paper we discuss the implementation of realistic clock noise and its synchronization via \ac{PTP} in \om.
The components presented in this paper are intended to assist engineers with the configuration of \ac{PTP} networks.
\end{abstract}

% \fi
%\IEEEpeerreviewmaketitle

% =======================================================================
%    Chapters
% =======================================================================

% =======================================================================
\section{Introduction}\label{sec:Intro}
% =======================================================================

\acp{RTS} are computer systems, where the progression of real time is of importance for the application which is carried out.
Distributed \acp{RTS} are a special class of systems: here, several distributed computer systems fulfill a common task which is linked to the progression of real-time.
For such an application, it is of utmost importance that each local node has reliable knowledge of the current time.
A typical example for such a system could be several robots that work together in a production line: it's easy to image that the correct operation of the overall system relies on the timely coordination between the individual systems.

Any physical clock device suffers from imperfections~\cite{ScienceOfTimekeeping}, and an unsynchronized ensemble of clocks could drift away from each other without an upper bound over time.
To construct a distributed global time base, different synchronization methods have been developed, each with unique benefits and drawbacks.
The \acf{PTP}, which is specified in IEEE~1588\cite{IEEE1588_2008}, presents another candidate to solve this problem.
\ac{PTP} promises sub-microsecond precision, while only requiring moderate costs.
The target domain for \ac{PTP} are local networks, where it is acceptable that each node provides hardware support for timestamping egressing and ingressing network frames.

\ac{PTP} presents many different options for its implementation and configuration.
To aid system designers with the task of \ac{DSE}, a simulation framework for \ac{PTP} is of great value.
For my master thesis~\cite{ThesisWoife}, I have developed such a simulation framework using \om\footnote{\url{https://omnetpp.org/}}.
This paper gives an overview over the \om specific aspects of this project.

% -----------------------------------------------------------------------
\subsection{Related Work}\label{sec:RelWork}
% -----------------------------------------------------------------------

This paper describes a subset of what I have been working of for my master thesis \cite{ThesisWoife}.
My thesis deals with the simulation of oscillator noise and the synchronization of oscillators via the \ac{PTP} which is specified in IEEE~1588.
An overall discussion of clocks and clock noise can be found in \cite{ScienceOfTimekeeping} and \cite{Handbook}.
An algorithm on how to simulate a particular important noise type (\acf{PLN}) is given in \cite{Kasdin92}.
My own noise simulation library is a modification for this approach to make it more suitable for \ac{DES} environments like \om.
In \cite{Gaderer11} an efficient implementation of the above mentioned noise simulation and its implementation in \om are described, however to the best knowledge of the author this implementation is not freely available and the description contains not enough details to reproduce their results.
Nevertheless this publication has served as an important guideline for the design of my own implementation.

The focus of this paper is on the \om-specific parts of my implementation.
Another publication named \emph{A Simulation Framework for IEEE~1588} will be presented at the 2016 International IEEE Symposium on Precision Clock Synchronization for Measurement, Control and Communication.
This other publication will contain a more detailed discussion of my noise simulation library.

% -----------------------------------------------------------------------
\subsection{Focus of this work}
% -----------------------------------------------------------------------

The goal for my thesis was to implement a simulation of clock synchronization via \ac{PTP}.
As I had to keep the project focus limited due to time constraints, I could only implement a subset of all components that would be present in a real physical \ac{PTP} network.
What I have selected for my simulation is basically a minimal set of components to get a meaningful \ac{PTP} simulation.
It consists of the following items:

\begin{itemize}
 \item \textbf{Noise generation:}

    \begin{itemize}
        \item A generic clock model in \om, which can be extended by various noise implementations
        \item A simulation library for \acl{PLN}, which is a noise type commonly found in oscillators
    \end{itemize}

 \item \textbf{Noise estimation:}

    \begin{itemize}

        \item A \ac{PTP} stack in \om, which implements most parts of the IEEE~1588-2008 standard, especially all clock and delay mechanism types
        \item An adaption of this generic \ac{PTP} stack to implement \ac{PTP} over \ether
        \item A generic model for \ac{PTP} network nodes in \om

    \end{itemize}

 \item \textbf{Noise cancellation:}

    \begin{itemize}
        \item A basic clock servo in \om based on a \ac{PI} controller
    \end{itemize}

\end{itemize}

\noindent All of the above mentioned components have been designed to be modular and extensible.

% -----------------------------------------------------------------------
\subsection{Structure of this work}
% -----------------------------------------------------------------------

In \autoref{sec:PTP}, we present a short overview of how \ac{PTP} works.
\autoref{sec:ClockModel} discusses clocks and clock noise, and how they are modeled in our simulation.
The structure of our \ac{PTP} implementation in \om is described in \autoref{sec:LibPTP}.
Finally, \autoref{sec:Conclusion} sums up the current project state, while \autoref{sec:FutureWork} gives an outlook to possible improvements.

% =======================================================================
\section{Precision Time Protocol}\label{sec:PTP}
% =======================================================================

This section intends to give a short overview over the principles of \ac{PTP}.
For a detailed discussion the interested reader is referred to the literature, e.g. \cite{BookEidson}.

% \emph{Remark:} Due to the limited space, the description of \ac{PTP} given here is focused on simple cases, and does not go into details.

% -----------------------------------------------------------------------
\subsection{Principal of operation}
% -----------------------------------------------------------------------

In a \ac{PTP} network, the individual nodes will dynamically establishes a hierarchy with master-slave relationships.
Two adjacent master and slave nodes then exchange information to allow the slave the estimation of its own clock offset relative to the master.
Using this information, a slave node can then modify its own local clock device to minimize its offset.

% -----------------------------------------------------------------------
\subsection{Overview}
% -----------------------------------------------------------------------

The overall functionality of \ac{PTP} consists of:

\begin{description}
 \item [Establishment of a clock hierarchy]
    \ac{PTP} nodes exchange the attributes of their clock with their neighboring nodes.
    This information is then used by a distributed algorithm (the \ac{BMC} algorithm), to establish a loop free master-slave hierarchy on the network.

 \item [Distribution of time information]
    \ac{PTP} nodes with ports in the master state will periodically publish their current time value, using so called \sync messages.

 \item [Path delay estimation]
    When a \ac{PTP} node receives a \sync message on a slave port, it needs to estimate how long this message has been on its way.
    To fulfill this task, \ac{PTP} specifies two methods for path delay estimation.

 \item [Configuration interface]
    Information of \ac{PTP} nodes is organized in standardized data sets.
    \ac{PTP} also specifies a management interface for accessing these data sets.

\end{description}

% -----------------------------------------------------------------------
\subsection{Information exchange}
% -----------------------------------------------------------------------

\begin{figure}[t!hbp]
  \centering
  \includegraphics[width=0.45\textwidth]{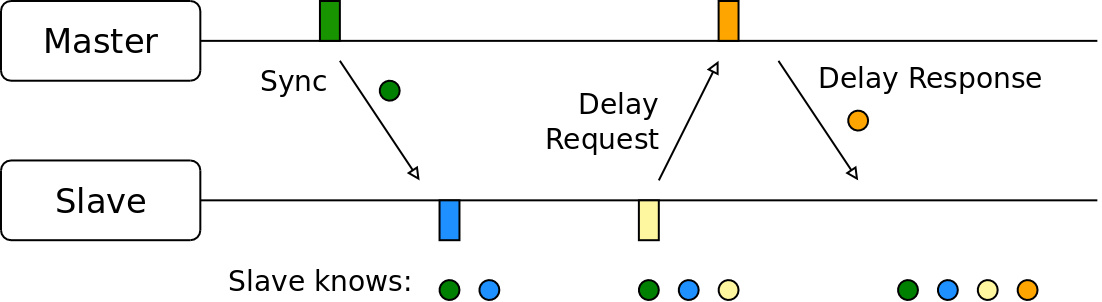}
  \caption{Synchronization principle of \ac{PTP}.}
  \label{fig:SyncPrinciple}
\end{figure}

An example for a possible communication between a \ac{PTP} master and a \ac{PTP} slave is shown in \autoref{fig:SyncPrinciple}.
The master periodically sends information about its current local time to the slave via \sync messages.
To estimate the time interval that \sync messages need until they arrive at the slave, the slave has to estimate the path delay.
It does so by periodically sending \dreq messages to the master, who then responds with \dresp messages.
The colored rectangles in the figure show the points in time when the frames are sent or received.
The colored circles below the figure sketch the knowledge on the slave side.
Using the collected information the slave is then able to estimate its own offset compared to the master.

% =======================================================================
\section{Clock Model}\label{sec:ClockModel}
% =======================================================================

In this section we discuss our approach for having realistic clock noise in our \om simulation model.

% -----------------------------------------------------------------------
\subsection{Problem Statement}
% -----------------------------------------------------------------------

The systems we would like to simulate are distributed networks, where each network node has access to a local clock.
We are interested in high precision clock synchronization, thus it is important for us that the individual clocks do not provide the exact value of real time, but only a local estimation.

The individual nodes rely on their clocks for two tasks:

\begin{itemize}
 \item \textbf{Timestamping:} Reading the current value of the local clock to generate a timestamp for a local event.
 \item \textbf{Scheduling:} Registering an event together with a future timestamp at the local clock.
    The event will be delivered when the timestamp is reached on the local clock.
\end{itemize}

\noindent \om provides \acp{API} for these two tasks with respect to real-time (\texttt{simTime()} and \texttt{scheduleAt()}).
We need to implement similar \acp{API} for the simulation of (noisy) local clocks.

% -----------------------------------------------------------------------
\subsection{Clock Model}
% -----------------------------------------------------------------------

In \cite{ScienceOfTimekeeping} a model for a digital clock is given.
This model consists of an oscillating device and a digital counter, which counts the oscillations.
The simulation model for our clocks is based on this model.
In a real clock, each of the two components could introduce noise.
For our simulation, we are not interested in the exact location inside a clock where the noise is introduced, but only on the effect that can be observed on the outside.
Thus, for our own model, we have modified the model from \cite{ScienceOfTimekeeping} as follows:

\begin{itemize}
 \item Both the oscillator and the counter are assumed to be absolutely \emph{perfect}
 \item In between these two we introduce a new component, called the noise generator.
    As the name already implies, its sole purpose is to introduce noise in our clock model.
 \item Finally, we have added a configurable scaling stage after the counter.
    This component can be used to synchronize the local clock to a reference clock using linear scaling.

\end{itemize}

\begin{figure}[t!hbp]
  \centering
  \includegraphics[width=0.45\textwidth]{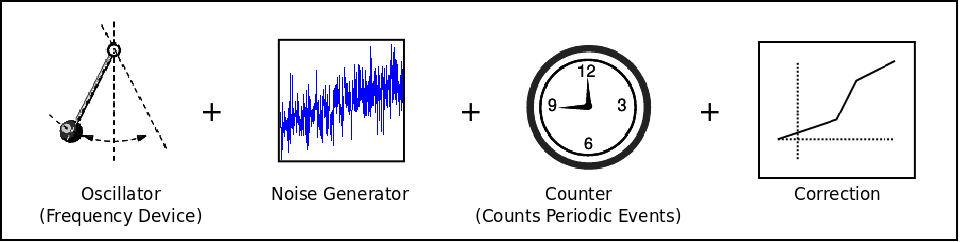}
  \caption{Clock model used in our simulation. The image is based on picture given in \cite{ScienceOfTimekeeping}.}
  \label{fig:ClockModel}
\end{figure}

A sketch of the complete clock model is given in \autoref{fig:ClockModel}.
\autoref{sec:ClockImpl} describes our implementation of this model.

% -----------------------------------------------------------------------
\subsection{Clock Noise}
% -----------------------------------------------------------------------

Clock devices suffer from various influences that disturb their correct operation.
These influences can be categorized as follows \cite{ScienceOfTimekeeping, Handbook}:

\begin{itemize}
 \item Systematic influences (temperature dependence, aging, frequency drift)
 \item Stochastic noise
\end{itemize}

\noindent The stochastic influences in most frequency sources (including quartz oscillators) can be modeled as a combination of noise processes where the power spectral density is related to the frequency as $S_y(f) \propto f^\alpha$.
This type of noise is referred to as \acf{PLN}.
For \ac{PLN} found in common oscillators, it turns out that $\alpha$ has integer values in the interval $-2 \knuthinterval 2$.

While it would be desirable to have all a noise model for our simulations that covers all these effects, I could only implement a subset of them because of timely constraints.
Depending on what kind of time intervals are most important for an application, different clock influences have to be considered or can be neglected.
As we want to simulate high precision synchronization, I have considered it valuable to have an implementation of \ac{PLN}, as it provides high-frequency noise as it is common in quartz oscillators.
On the other hand, influences like aging can probable neglected without affecting the overall simulation results too much.

I have designed and implemented a library for the simulation of \ac{PLN}, call \lpln \footnote{\url{https://github.com/w-wallner/libPLN}}.
The actual \ac{PLN} simulation is based on the approach by Kasdin and Walter \cite{Kasdin92}, and ideas from \cite{Handbook} and \cite{Gaderer11}.

% -----------------------------------------------------------------------
\subsection{Implementation}\label{sec:ClockImpl}
% -----------------------------------------------------------------------

\begin{figure}[t!hbp]
  \centering
  \includegraphics[width=0.3\textwidth]{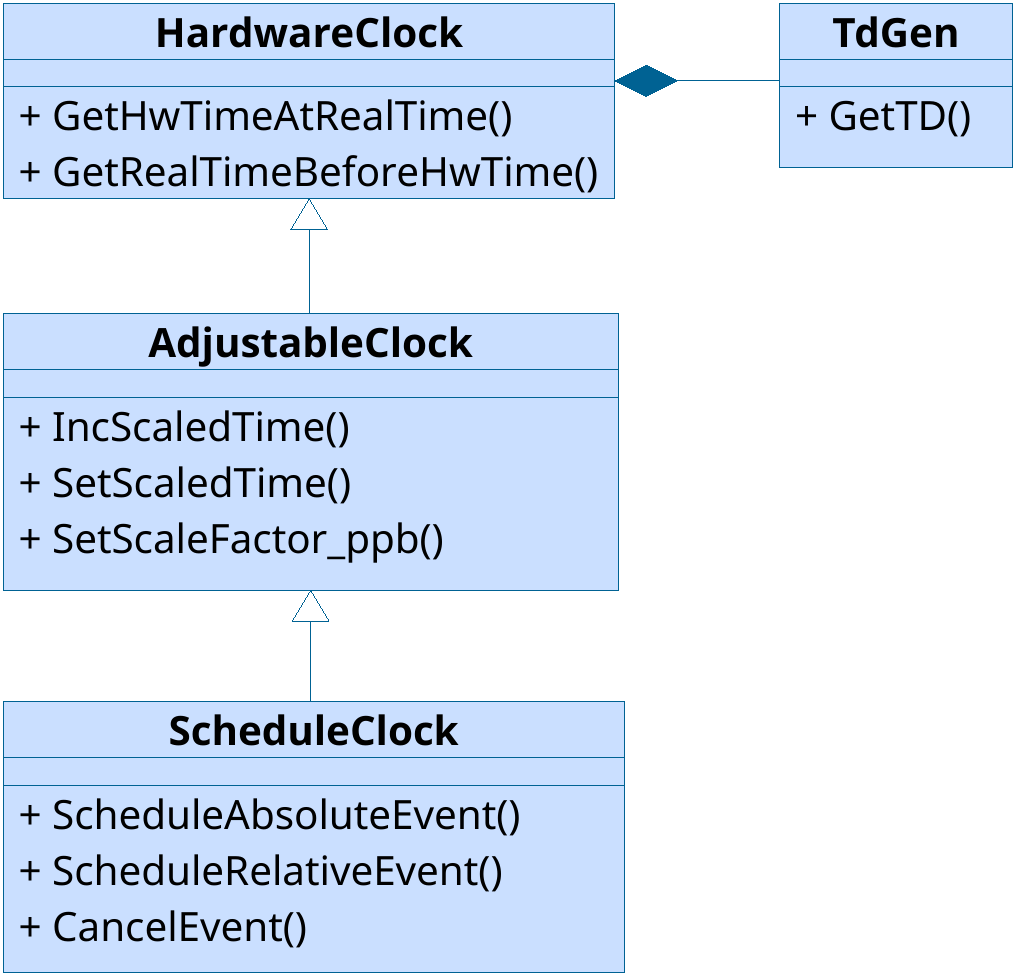}
  \caption{Hierarchy of our clock components}
  \label{fig:ClockHierarchy}
\end{figure}

To reduce the complexity of our implementation, the individual clock tasks are implemented in different classes.
\autoref{fig:ClockHierarchy} shows the relationship between the individual components of our model.

% ~~~~~~~~~~~~~~~~~~~~~~~~~~~~~~~~~~~~~~~~~~~~~~~~~
\paragraph{Hardware Clock}
% ~~~~~~~~~~~~~~~~~~~~~~~~~~~~~~~~~~~~~~~~~~~~~~~~~

The most basic class is the \texttt{HwClock}.
% This component corresponds to the quartz oscillator in a real system.
Its purpose is the translation between the continuous and perfect real-time, and a discrete imperfect local time estimate.
A hardware clock can not be adjusted, and its counter has no relationship to that of other clocks.

To fulfill its task, the hardware clock class relies on a component called \texttt{TdGen}, the \ac{TD} generator.
This component implements the current deviation of the local oscillator from the perfect real-time.
It is this component that contains the noise model used in the simulation.
In my implementation, it uses the above described \lpln and only simulates \acl{PLN}.
Further s to the noise model could be implemented in this component (e.g. temperature dependence).

The local time estimate $t_{est}$ for the real-time $t$ can then be written as $t_{est} = t + TD(t)$.
It is important that we can not assume to have an inverse for this function.
But we can rely on the fact that the local time estimation is always increasing.
This is important when we want to calculate the real-time when a specified local time estimate will be reached, as we need to this later to implement scheduling.

% ~~~~~~~~~~~~~~~~~~~~~~~~~~~~~~~~~~~~~~~~~~~~~~~~~
\paragraph{Adjustable Clock}
% ~~~~~~~~~~~~~~~~~~~~~~~~~~~~~~~~~~~~~~~~~~~~~~~~~

Adjustable clocks provide an abstraction on top of hardware clocks.
Their counter values can be modified, and their progress can be speed up or slowed down.
These features enable us to synchronize multiple adjustable clocks to each other.

% ~~~~~~~~~~~~~~~~~~~~~~~~~~~~~~~~~~~~~~~~~~~~~~~~~
\paragraph{Schedule Clock}
% ~~~~~~~~~~~~~~~~~~~~~~~~~~~~~~~~~~~~~~~~~~~~~~~~~

The finale clock class is the \texttt{ScheduleClock}, which provides an interface for other \om components to schedule future events.
Internally, it stores the scheduled events in an ordered list, and only the first one is actually scheduled.
For the scheduling, it relies on the \ac{API} provided by the hardware clock for real-time estimation together with \om's \texttt{scheduleAt()} functionality.

% =======================================================================
\section{PTP Implementation}\label{sec:LibPTP}
% =======================================================================

This section gives a brief overview over \lptp, our \om-based implementation of \ac{PTP}.

% -----------------------------------------------------------------------
\subsection{Node Architecture}
% -----------------------------------------------------------------------

\begin{figure}[t!hbp]
  \centering
  \includegraphics[width=0.35\textwidth]{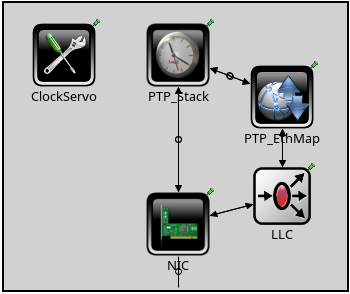}
  \caption{Architecture of a PTP Basic Node}
  \label{fig:PTP_Node}
\end{figure}

\ac{PTP} defines several different types of network nodes (\acp{OC}, \acp{BC} and \acp{TC}), as well as a large number of configuration options.
In \lptp, all these nodes are derived from a common base class, called \texttt{PTP\_BasicNode}.
The architecture of this base class is shown in \autoref{fig:PTP_Node}.
Most components from LibPTP are based on standard models from the INET library\footnote{\url{https://inet.omnetpp.org}}.

The architecture shown here is a design choice with the intention to represent a basic \ac{PTP}-capable network device.
Any hardware related components have been placed on our simulated \ac{NIC}.
Components outside the \ac{NIC} represent components that would typically be present as software in a real system.
The node model here is loosely based on the architecture of a PC with a \ac{PTP}-capable \ac{NIC}.
This is also why e.g. the clock that is used for the synchronization is shown as part of the simulated \ac{NIC}.

The layout of our node model is one of the topics where community feedback for the improvement of \lptp would be greatly appreciated.

% ~~~~~~~~~~~~~~~~~~~~~~~~~~~~~~~~~~~~~~~~~~~~~~~~~
\paragraph{PTP Stack}
% ~~~~~~~~~~~~~~~~~~~~~~~~~~~~~~~~~~~~~~~~~~~~~~~~~

The most important component is the \texttt{PTP\_Stack}, which is our implementation of the protocol as it is specified in IEEE~1588-2008 \cite{IEEE1588_2008}.

% ~~~~~~~~~~~~~~~~~~~~~~~~~~~~~~~~~~~~~~~~~~~~~~~~~
\paragraph{PTP Ethernet Mapping}
% ~~~~~~~~~~~~~~~~~~~~~~~~~~~~~~~~~~~~~~~~~~~~~~~~~
\ac{PTP} does not depend on a single lower layer technology, and various mappings are standardized, e.g for \ac{PTP} over \ac{UDP}, plain \ether, or different fieldbus systems.
For our project, we focused on \ac{PTP} over \ether, which is specified in Annex~F of IEEE~1588-2008.
The component \texttt{PTP\_EthMap} implements this mapping.

% ~~~~~~~~~~~~~~~~~~~~~~~~~~~~~~~~~~~~~~~~~~~~~~~~~
\paragraph{Clock Servo}
% ~~~~~~~~~~~~~~~~~~~~~~~~~~~~~~~~~~~~~~~~~~~~~~~~~

\ac{PTP} only specifies a way to estimate the offset of a slave clock in reference to a master clock.
How this offset is minimized is the responsibility of the individual implementations, in particular that of the implemented clock servo controller.
Our simulation provides a generic interface for clock servos, and an example implementation using a \ac{PI}-based clock servo design.

% ~~~~~~~~~~~~~~~~~~~~~~~~~~~~~~~~~~~~~~~~~~~~~~~~~
\paragraph{LLC}
% ~~~~~~~~~~~~~~~~~~~~~~~~~~~~~~~~~~~~~~~~~~~~~~~~~

The component labeled \texttt{LLC} in \autoref{fig:PTP_Node} is the \ac{LLC}.
It provides access to the \ac{NIC} based on the EtherType field of \ether frames.

% ~~~~~~~~~~~~~~~~~~~~~~~~~~~~~~~~~~~~~~~~~~~~~~~~~
\paragraph{NIC}
% ~~~~~~~~~~~~~~~~~~~~~~~~~~~~~~~~~~~~~~~~~~~~~~~~~

\begin{figure}[t!hbp]
  \centering
  \includegraphics[width=0.35\textwidth]{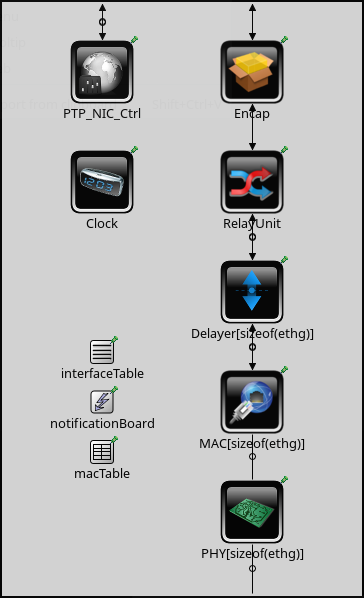}
  \caption{Internal architecture of our \ac{PTP}-capable \ac{NIC}}
  \label{fig:PTP_NIC}
\end{figure}

Each \ac{PTP} node has a \ac{NIC} with \ac{PTP} hardware support.
\autoref{fig:PTP_NIC} shows the internal architecture of our \ac{NIC} model.

% ~~~~~~~~~~~~~~~~~~~~~~~~~~~~~~~~~~~~~~~~~~~~~~~~~
\paragraph{Clock}
% ~~~~~~~~~~~~~~~~~~~~~~~~~~~~~~~~~~~~~~~~~~~~~~~~~

The clock inside the \ac{NIC} implements the clock model described in \autoref{sec:ClockModel}.

% ~~~~~~~~~~~~~~~~~~~~~~~~~~~~~~~~~~~~~~~~~~~~~~~~~
\paragraph{Relay Unit}
% ~~~~~~~~~~~~~~~~~~~~~~~~~~~~~~~~~~~~~~~~~~~~~~~~~

In case a network node with more than one physical network interface is simulated, it is the task of the \texttt{RelayUnit} component to distributed ingressing and egressing frames to the correct ports.

% ~~~~~~~~~~~~~~~~~~~~~~~~~~~~~~~~~~~~~~~~~~~~~~~~~
\paragraph{Delay}
% ~~~~~~~~~~~~~~~~~~~~~~~~~~~~~~~~~~~~~~~~~~~~~~~~~

To simulate arbitrary scheduling/queuing delays, a special \texttt{Delayer} component is added to our simulation model.

% ~~~~~~~~~~~~~~~~~~~~~~~~~~~~~~~~~~~~~~~~~~~~~~~~~
\paragraph{MAC}
% ~~~~~~~~~~~~~~~~~~~~~~~~~~~~~~~~~~~~~~~~~~~~~~~~~

The \ac{MAC} is based on the \texttt{EtherMAC} from INET, but adds timestamping support for \ac{PTP} frames.

% ~~~~~~~~~~~~~~~~~~~~~~~~~~~~~~~~~~~~~~~~~~~~~~~~~
\paragraph{PHY}
% ~~~~~~~~~~~~~~~~~~~~~~~~~~~~~~~~~~~~~~~~~~~~~~~~~

The \ac{PHY} model provides support for simulating asymmetric paths.

% -----------------------------------------------------------------------
\subsection{Simulations}
% -----------------------------------------------------------------------

Using the combination of \lptp together with \lpln allows us to carry out simulations of clock synchronization via \ac{PTP} in \om.

% ~~~~~~~~~~~~~~~~~~~~~~~~~~~~~~~~~~~~~~~~~~~~~~~~~
\paragraph{Sync interval}
% ~~~~~~~~~~~~~~~~~~~~~~~~~~~~~~~~~~~~~~~~~~~~~~~~~

On of the most important parameters for a \ac{PTP} configuration is the interval for the sending of \sync messages.
The synchronization interval in \ac{PTP} is given via the \texttt{logSyncInterval} parameter as $T_{sync} = 2^{logSyncInterval}$, measured in seconds.
Configuring a long interval might degrade the reachable clock synchronization.
On the other hand, if an interval is configured which is too short, we might waste network bandwidth without any benefit for the precision of our global time-base.
Using \om's support for parameter studies, we can configure a \ac{PTP} network and find out an optimal value empirically.

\begin{figure}[t!hbp]
  \centering
  \includegraphics[width=0.5\textwidth]{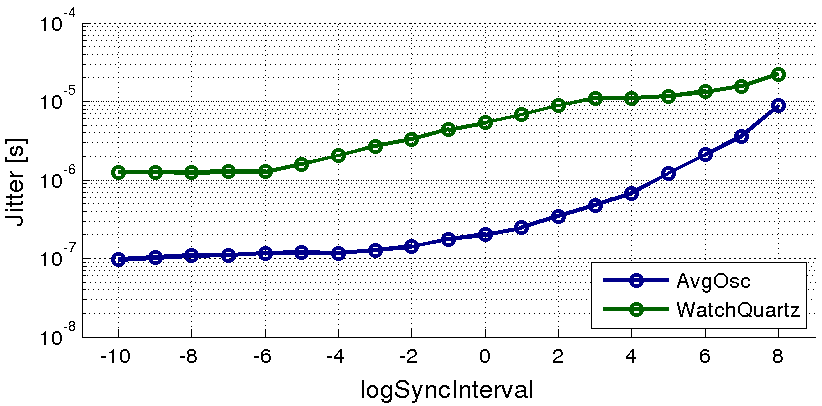}
  \caption{Parameter study for showing the effect of different synchronization intervals.}
  \label{fig:SyncInterval_Jitter}
\end{figure}

Suppose we have a simple network, consisting of only two \ac{PTP} nodes.
One of them has an excellent clock, and is the \ac{PTP} master, while the other has a cheap oscillator and is the \ac{PTP} slave.
Carrying out a parameter study for possible synchronization interval configurations could lead to a graph as shown in \autoref{fig:SyncInterval_Jitter}.
The image shows the worst-case jitter of the slave's time estimate relative to the time of the master.

The slave was simulated to have one of two oscillators: either a cheap quartz oscillator as it could be found in typical consumer electronic devices or a quartz typical for wrist watches.
It can be seen that for both types of oscillators, the jitter decreases for shorter intervals, until it reaches an oscillator specific lower bound.
Further decreasing the interval won't lead to better clock synchronization.

% ~~~~~~~~~~~~~~~~~~~~~~~~~~~~~~~~~~~~~~~~~~~~~~~~~
\paragraph{Tracing and debugging}
% ~~~~~~~~~~~~~~~~~~~~~~~~~~~~~~~~~~~~~~~~~~~~~~~~~

\begin{figure}[t!hbp]
  \centering
  \includegraphics[width=0.5\textwidth]{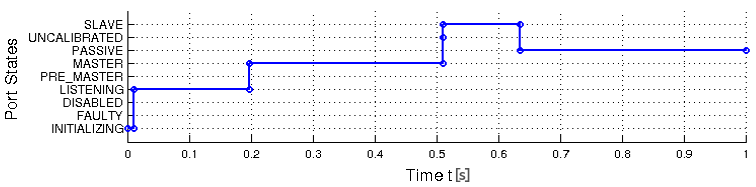}
  \caption{Trace of a \ac{PTP} port state}
  \label{fig:PortStates}
\end{figure}

\lptp provides also sophisticated tracing and debugging support.
All relevant \ac{PTP} properties are available as signals and statistics.
As an example, a trace of a \ac{PTP} port state is shown in \autoref{fig:PortStates}.

\noindent Additionally, it is possible to enable various logging messages.
For example, it is easily possible to get a complete trace output of every single decision that is carried out when the nodes execute the \ac{BMC} or the \ac{DSC} algorithms.

% =======================================================================
\section{Conclusion}\label{sec:Conclusion}
% =======================================================================

Our project has shown that it is feasible to implement realistic clock noise in \om, as well as to carry out simulations of clock synchronization via \ac{PTP}.
To best knowledge of the author, it is the most sophisticated implementation of IEEE~1588 in \om which is freely available.
All source code is released under open source licenses\footnote{Most parts are released under the GPL, the rest under the BSD license.} on Github\footnote{\url{https://github.com/w-wallner/libPTP}}.
Additionally, also a portable library called \lpln for the efficient simulation of \acl{PLN} in \ac{DES}-environment like \om was implemented.
Again, everything is available to the community as open source on Github\footnote{\url{https://github.com/w-wallner/libPTL}}.
The components presented here should help engineers with the difficult task of \ac{PTP} \ac{DSE}.

All of the implemented components have been designed to be extensible, so that other users can modify and improve them for their own projects.

As a side effect, having an \om-based implementation of \ac{PTP} has proven to be an excellent tool when teaching other people how \ac{PTP} works.

% =======================================================================
\section{Future Work}\label{sec:FutureWork}
% =======================================================================

I have worked on this project mainly for my master thesis.
During this time, I have worked on it in private.
As my thesis is finished now, I have released all related source code under open source licenses, in the hope to find interested community members for further enhancements.

Future improvements to \lptp could include:

\begin{itemize}

\item \textbf{Extending PTP options:}
Currently only a subset of the standardized \ac{PTP} options is implemented in \lptp.
Extending this support would further enhance the capabilities for \ac{DSE}.
Of interest would be e.g. unicast support, \ac{PTP} over \ac{UDP} or support for different \ac{PTP} profiles.

\item \textbf{Mainlining LibPTP:}
\lptp is derived from components in the INET library, but it is not related to upstream development.
This implies that the projects will drift further apart.
It would be of great benefit to try to merge at least some of the work upstream.

\item \textbf{Improving the noise model:}
Currently, only stochastic noise of oscillators is modeled (using \lpln).
It would be interesting to add support for deterministic influences like temperature to the simulation model.
This would allow the simulation to provide even more realistic results.

\end{itemize}

% =======================================================================
\section{Acknowledgements}\label{sec:Ack}
% =======================================================================

I would like to thank my supervisor Armin Wasicek for his support during my master thesis.
I am grateful to both the \om and INET communities: without the availability of their excellent tools/libraries, my \ac{PTP} implementation would not have been possible.

% =======================================================================
%  Acronyms
% =======================================================================

\begin{acronym}[OMNeT++]            % This Option sets the space between acronym and description
\setlength{\itemsep}{-\parsep}      % This is for a more compact style (less space between items)
%-------------------------------------------------------------------------------------------------------
\acro   {ADEV}      {Allan Deviation}
\acro   {API}       {Application Programming Interface}
\acro   {AVAR}      {Allan Variance}
%-------------------------------------------------------------------------------------------------------
\acro   {BC}        {Boundary Clock}
\acro   {BMC}       {Best Master Clock}
%-------------------------------------------------------------------------------------------------------
\acro   {COTS}      {Commercial off-the-shelf}
\acro   {CM}        {Compound Module}
\acro   {CSV}       {Comma-separated values}
%-------------------------------------------------------------------------------------------------------
\acro   {DSC}       {Data Set Comparision}
\acro   {DES}       {Discrete Event Simulation}
\acro   {DSE}       {Design Space Exploration}
%-------------------------------------------------------------------------------------------------------
\acro   {E2E}       {End-to-End}
%-------------------------------------------------------------------------------------------------------
\acro   {FFM}       {Flicker Frequency Modulation}
\acro   {FFD}       {Fractional Frequency Deviation}
\acro   {FIR}       {Finite Impulse Response}
\acro   {FPM}       {Flicker Phase Modulation}
\acro   {FSA}       {Frequency Stability Analysis}
\acro   {FFT}       {Fast Fourier transform}
\acro   {FFTW}      {Fastest Fourier Transform in the West}
%-------------------------------------------------------------------------------------------------------
\acro   {GPL}       {GNU General Public License}
\acro   {GPS}       {Global Positioning System}
\acro   {GSL}       {GNU Scientific Library}
\acro   {GUI}       {Graphical User Interface}
%-------------------------------------------------------------------------------------------------------
\acro   {HP}        {High Pass}
%-------------------------------------------------------------------------------------------------------
% \acro   {IA}        {Industrial Automation}
\acro   {IDE}       {Integrated Development Environment}
\acro   {IEEE}      {Institute of Electrical and Electronics Engineers}
\acro   {IIR}       {Infinite Impulse Response}
\acro   {ISPCS}     {International IEEE Symposium on Precision Clock Synchronization for Measurement, Control and Communication}
%-------------------------------------------------------------------------------------------------------
\acro   {LAN}       {Local Area Network}
\acro   {LLC}       {Logical Link Control}
%-------------------------------------------------------------------------------------------------------
\acro   {MAC}       {Media Access Control}
\acro   {MAVAR}     {Modified Allan Variance}
%-------------------------------------------------------------------------------------------------------
\acro   {NED}       {Network Description}
\acro   {NIC}       {Network Interface Card}
\acro   {NTP}       {Network Time Protocol}
%-------------------------------------------------------------------------------------------------------
\acro   {OC}        {Ordinary Clock}
\acro   {OCXO}      {Oven Controlled Crystal Oscillator}
\acro   {OMNeT++}   {Objective Modular Network Testbed in C++}
\acro   {OS}        {Operating System}
%-------------------------------------------------------------------------------------------------------
\acro   {P2P}       {Peer-to-Peer}
\acro   {PC}        {Personal computer}
\acro   {PHY}       {physical layer}
\acro   {PI}        {pro\-por\-tional-integral}
\acro   {PID}       {proportional-integral-derivative}
\acro   {PLN}       {Powerlaw Noise}
\acro   {PSD}       {Power Spectral Density}
\acro   {PTP}       {Precision Time Protocol}
%-------------------------------------------------------------------------------------------------------
\acro   {RGMII}     {Reduced Gigabit Media-Independent Interface}
\acro   {RTS}       {Real-Time System}
\acro   {RW}        {Random Walk}
%-------------------------------------------------------------------------------------------------------
\acro   {SP}        {Scale Point}
\acro   {SM}        {Simple Module}
%-------------------------------------------------------------------------------------------------------
\acro   {TC}        {Transparent Clock}
\acro   {TD}        {Time Deviation}
\acro   {TDG}       {\ac{TD} Generator}
\acro   {TDO}       {\ac{TD} Oracle}
\acro   {TDV}       {\ac{TD} Vector}
\acro   {TDVG}      {\ac{TDV} Generator}
\acro   {TDVS}      {\ac{TDV} Storage}
\acro   {TDEC}      {\ac{TDE} Chain}
\acro   {TDE}       {\ac{TD} Estimator}
\acro   {TSA}       {Time Series Analysis}
%-------------------------------------------------------------------------------------------------------
\acro   {UML}       {Unified Modeling Language}
\acro   {UDP}       {User Datagram Protocol}
%-------------------------------------------------------------------------------------------------------
\acro   {WFM}       {White Frequency Modulation}
\acro   {WPM}       {White Phase Modulation}
%-------------------------------------------------------------------------------------------------------
\end{acronym}
% =======================================================================
%    REFERENCES
% =======================================================================



%\bibliographystyle{IEEEtran}

\end{document}